\documentclass[11pt]{article}
\usepackage{amsmath}
\usepackage{epsfig}

\begin{document}

\begin{center}
\bigskip

{\Large Long memory stochastic volatility in option pricing}

\bigskip

{\large Sergei Fedotov and Abby Tan}

\bigskip

Department of Mathematics, UMIST, M60 1QD UK

\bigskip

Submitted to IJTAF on 16 March 2004

\bigskip

\textbf{Abstract}

\bigskip
\end{center}

The aim of this paper is to present a stochastic model that accounts for the
effects of a long-memory in volatility on option pricing. The starting point
is the stochastic Black-Scholes equation involving volatility with
long-range dependence. We define the stochastic option price as a sum of
classical Black-Scholes price and random deviation describing the risk from
the random volatility. By using the fact that the option price and random
volatility change on different time scales, we derive the asymptotic
equation for this deviation involving fractional Brownian motion. The
solution to this equation allows us to find the pricing bands for options.

Keywords: Long memory, stochastic volatility, option pricing

\bigskip

\section{ Introduction}

Over the last few years, self-similarity and long-range dependence have
become important concepts in analyzing the financial time series \cite
{Peters,man}. There is strong evidence that the return, $r_{t},$ has little
or no autocorrelation, whereas its square, $r_{t}^{2},$ or absolute return, $%
|r_{t}|,$ exhibit noticeable autocorrelation \cite{Camp}. This phenomenon
can be described by the ARCH(p) model \cite{Engle1982} or its GARCH(p,q)
extension \cite{Bollerslev1986}. However, the exponential decay for $\lambda
_{s}=cov(r_{t}^{2},r_{t+s}^{2})$ is believed to be too fast to describe
correctly the persistent dependence between the series observations as the
time lag increases. It turns out \cite{Poon,Beran} that the models with
hyperbolic decay which have slowly decaying covariances provide better
fitting to financial time series. The characteristic feature of these models
is that their covariance $\lambda _{s}$ has the power law decay $s^{2d-1}$ ($%
0<d<1/2)$ for the large lag $s$ such that
\begin{equation}
\sum\limits_{s=0}^{\infty }\lambda _{s}=\infty \text{ }.  \label{longmemory}
\end{equation}
The financial series is said to have \textit{long memory} if it
displays this property \cite{Beran}. This means that the square of
returns that are far apart are strongly correlated, since the
correlations decay
very slowly to zero. Let us note that the series is said to have \textit{%
short memory} if the autocovarience is summable. Evidences for \textit{long
memory} in daily absolute or squared returns have been well documented (see,
for example, \cite{Ding,Granger,F}). The review of \textit{long memory}
models leading to (\ref{longmemory}) can be found in \cite{Beran}.

The natural question arises: what is the implication of this long-range
dependence in volatility on the option pricing. It is well known that the
existence of `smile' and `frown' in the implied volatility graph contradicts
the Black-Scholes assumption of constant volatility. To remedy this
shortcoming, many stochastic volatility models have been proposed (see, for
example, \cite{Sircar,Lewis,Taylor}). Despite the large volume of literature
discussing stochastic volatility, the effects of long-memory on option
pricing are relatively unexplored. Several attempts have been made to
understand the role of long-range dependence in volatility on derivative
pricing. Comte and his colleagues generalized the classical Heston model to
account for long memory features of stochastic volatility \cite
{Comte1,Comte2}. The main idea was to define the volatility process as the
fractional integration of the short memory process. The main difficulty with
this approach is that the volatility is not a directly tradeable asset, and
therefore the classical Black-Scholes hedging can not be applied. It leads
to the appearance of an unknown parameter, namely, the market price of
volatility risk \cite{Sircar,Lewis}. The problem is that this parameter is
not directly observable and one has to make additional assumptions regarding
the pricing of volatility risk. An extension of the Stein and Stein model
for a long memory volatility case was given in \cite{Hu2}, where the option
price was obtained through a risk-minimization procedure (see also other
works \cite{Taylor2,Dj}). A different approach to option pricing with
stochastic volatility has been suggested in \cite{Sircar2}. Instead of
finding the exact option price the authors focused on the pricing bands for
options that account for the random volatility risk. However, they
considered only weakly correlated random volatility and did not discuss the
long memory effects. Stochastic volatility can be also treated by a
stochastic optimization approach based on a risk minimization procedure \cite
{FM}.

It is the purpose of this paper to present a simple stochastic volatility
model that accounts for the long memory effects of stochastic volatility on
option pricing. We start with the classical Black-Scholes equation with
random volatility which has long-range dependence. The aim is to find the
option price as a sum of Black-Scholes price with average volatility and
random deviation describing the risk from random volatility. A main feature
of this approach is that it does not need an estimation of the market price
volatility risk \cite{Sircar2}.

It should be noted that there exists a family of continuous random processes
which capture the long memory property including the well-known fractional
Brownian motion (fBm) \cite{Comte}. A substantial amount of research has
been done in an attempt to model the stock return (see, for example, \cite
{Hu} and the references therein). In particular, it has been shown that the
simple replacement of the Wiener process by fBm leads to arbitrage
opportunities in the market \cite{Rogers}. Recall that the fractional
Brownian motion with Hurst coefficient $H$ is the Gaussian process $B_{d}(t)$
with mean $E\left\{ B_{d}(t)\right\} =0$ and covariance \cite{Comte}
\begin{equation}
E\left\{ B_{d}(t)B_{d}(s)\right\} =\frac{1}{2}%
(t^{2d+1}+s^{2d+1}-|t-s|^{2d+1}),  \label{Fractional}
\end{equation}
where $d$ is the difference parameter: $d=H-1/2,$ and $E$ denotes the
expectation. In particular, it follows from (\ref{Fractional}) that $%
E\{B_{d}^{2}\}=t^{1+2d}$. If $H=1/2$ $(d=0)$, then $B_{d}(t)$ is identical
to the standard Brownian motion $B(t).$ If $1/2<H<1$ $(0<d<1/2),$ then $%
B_{d}(t)$ has a long-range dependence in the sense that the sequence of
increments $g_{d}(n)=B_{d}(n)-$ $B_{d}(n-1)$ is strongly correlated, namely,
$E\{g_{d}(n)g_{d}(n+k)\}\sim k^{2d-1}$ as $k\rightarrow \infty .$ The
stochastic calculus for fractional Brownian motion for the Hurst exponent in
the interval $(1/2,1)$ is given in \cite{Duncan}

\section{Incorporation of long memory volatility into the option pricing
model}

We start with the Black-Scholes equation with the random volatility $\sigma
(t),$
\begin{equation}
\frac{\partial C}{\partial t}+\frac{1}{2}\sigma ^{2}(t)S^{2}\frac{\partial
^{2}C}{\partial S^{2}}+rS\frac{\partial C}{\partial S}-rC=0,  \label{BS}
\end{equation}
where $C$ is the call option price, $r$ is the interest rate, and $t$ is the
time ($0\leq t\leq T$). We adopt the idea that the volatility's time
variations are small compared with time scale of the option \cite{Sircar2}.
We assume that $\sigma ^{2}(t)$ is the stationary long memory process with
the following statistical characteristics:
\begin{equation}
E\{\sigma ^{2}(t)\}=\overline{\sigma ^{2}},\;\;\;E\{(\sigma ^{2}(t)-%
\overline{\sigma ^{2}})(\sigma ^{2}(t+s)-\overline{\sigma ^{2}})\}=g(s)=%
\frac{c}{(s_{0}+s)^{\alpha }},  \label{covar1}
\end{equation}
where $E\{\cdot \}$ is the expectation, $g(s)$ is the covariance, and the
parameter $\alpha $ obeys
\begin{equation}
0<\alpha <1.
\end{equation}
It follows from (\ref{covar1}) that for large lag $s,$ the covariance $g(s)$
decreases to zero like a power law $s^{-\alpha },$ while the spectrum $\rho
\left( \lambda \right) $ goes to infinity at the origin \cite{Beran}
\begin{equation}
\rho (\lambda )\sim \frac{c_{\rho }}{\lambda ^{1-\alpha }}\;\;\;\text{as}%
\;\;\;\lambda \rightarrow 0.
\end{equation}
The parameter $\alpha $ can be regarded as a measure of the intensity of the
long-range dependence of the volatility. A variety of techniques are
available to estimate the parameter $\alpha $ from financial time series (see%
\textit{\ }\cite{Beran})\textit{.} The characteristic feature of (\ref
{covar1}) is that the integral $\int_{0}^{t}g(s)ds$ diverges as $%
t\rightarrow \infty .$

We should remark that our assumption that the typical time scales of
volatility variations are small compared with the time scale of the
derivative contract is not inconsistent with the long memory properties of
the volatility (\ref{covar1}). A similar situation occurs in the
renormalization theory of turbulent diffusivity in which the random velocity
field with a power law spectrum involves a continuous range of time/space
scales and all of them are less than the integral scales (see\textit{\ }\cite
{Avellaneda}). If we assume that the time variable $t$ is measured in $t_{d}$
units (several days), then the spectrum $\rho _{\varepsilon }(\lambda )$ can
be written as follows
\begin{equation}
\rho _{\varepsilon }(\lambda )\sim \frac{c_{\rho }}{\lambda ^{1-\alpha }}%
\;\;\;\text{for}\;\;\;\varepsilon <\lambda <<1,\;\;\;\rho _{\varepsilon
}(\lambda )=0\;\;\;\text{for}\;\;\;0\leq \lambda \leq \varepsilon ,
\label{new2}
\end{equation}
where $\varepsilon $ is the small parameter
\begin{equation}
\varepsilon =\frac{t_{d}}{T}.
\end{equation}
The spectrum (\ref{new2}) describes the situation when there are no
time-scales in $\sigma ^{2}(t)$ greater than the expiry date $T.$ One can
also write this using the infrared cut-off function $\psi _{0}:$ the
spectrum $\rho _{\varepsilon }(\lambda )$ can be represented as $\rho
_{\varepsilon }(\lambda )\sim \frac{c_{\rho }}{\lambda ^{1-\alpha }}\psi
_{0}\left( \frac{\lambda }{\varepsilon }\right) $, where $\psi _{0}=1$ for $%
\lambda >\varepsilon $ and $\psi _{0}=0$ for $\lambda \leq \varepsilon .$
This idea was introduced by Avellaneda and Majda in \cite{Avellaneda}, where
they analyzed turbulent transport\ for a random velocity with a power law
spectrum. Of course, in the limit $\varepsilon \rightarrow 0$ the spectrum, $%
\rho _{\varepsilon }(\lambda )$ tends to $\rho (\lambda )$ which leads to
the infrared divergence of corresponding integrals. In what follows we
consider only the asymptotic regime $\varepsilon \rightarrow 0,$ therefore
one can use the idealized spectrum function $\rho (\lambda )$ instead of\ $%
\rho _{\varepsilon }(\lambda )$ with the infrared cut-off. The details
concerning the renormalization procedure, non-trivial scaling and infrared
divergent integrals can be found in \cite{Avellaneda}.

To deal with the forward problem we introduce the time-to-maturity
\begin{equation}
\tau =T-t  \label{time}
\end{equation}
and measure it in terms of one-year units. The stochastic volatility $\sigma
(\tau )$ becomes a rapidly fluctuating function. It is convenient to split $%
\sigma ^{2}(\tau )$ into the mean value, $\overline{\sigma ^{2}},$ and a
rapidly varying component, $v,$ as follows
\begin{equation}
\sigma ^{2}(\tau )=\overline{\sigma ^{2}}+v\left( \frac{\tau }{\varepsilon }%
\right) .  \label{split}
\end{equation}
It is evident that the autocorrelation function for $v$ is given by $g(s)$
(see (\ref{covar1})). Using (\ref{time}) and (\ref{split}), we find that the
corresponding stochastic option price $C^{\varepsilon }(\tau ,S)$ satisfies
the following stochastic PDE:
\begin{equation}
\frac{\partial C^{\varepsilon }}{\partial \tau }=\frac{1}{2}\overline{\sigma
^{2}}S^{2}\frac{\partial ^{2}C^{\varepsilon }}{\partial S^{2}}+\frac{1}{2}%
v\left( \frac{\tau }{\varepsilon }\right) S^{2}\frac{\partial
^{2}C^{\varepsilon }}{\partial S^{2}}+rS\frac{\partial C^{\varepsilon }}{%
\partial S}-rC^{\varepsilon }  \label{stochasticPDE}
\end{equation}
subject to the initial condition
\begin{equation}
C^{\varepsilon }(0,S)=\max (S-K,0),  \label{initial}
\end{equation}
where $K$ is the strike price.

Our purpose now is to analyze the asymptotic behavior of $C^{\varepsilon
}(\tau ,S)$ as $\varepsilon \rightarrow 0.$ We split the option price $%
C^{\varepsilon }(\tau ,S)$ into the sum of the deterministic price, $%
\overline{C}(\tau ,S),$ and the random deviation, $Z^{\varepsilon }(\tau
,S), $ with the anomalous scaling factor, $\varepsilon ^{1/2-d},$ to get
\begin{equation}
C^{\varepsilon }(\tau ,S)=\overline{C}(\tau ,S)+\varepsilon
^{1/2-d}Z^{\varepsilon }(\tau ,S),  \label{split2}
\end{equation}
where $\overline{C}(\tau ,S)$ is the classical Black-Scholes price, which
satisfies \cite{Hull,Wilmott}
\begin{equation}
\frac{\partial \overline{C}}{\partial \tau }=\frac{1}{2}\overline{\sigma ^{2}%
}S^{2}\frac{\partial ^{2}\overline{C}}{\partial S^{2}}+rS\frac{\partial
\overline{C}}{\partial S}-r\overline{C}.  \label{classical}
\end{equation}
Note that the case $d=0$ corresponds to the short memory volatility model
considered in \cite{Sircar2}. In this paper we assume that the parameter $d$
varies in the range:
\begin{equation}
0<d<\frac{1}{2}.  \label{range}
\end{equation}
The expression for the parameter $d$ in terms of $\alpha $ is derived in
Appendix A (see (A-4)):
\begin{equation}
d=\frac{1-\alpha }{2}.
\end{equation}
Substituting (\ref{split2}) into (\ref{stochasticPDE}) and using (\ref
{classical}), we get the equation for $Z^{\varepsilon }(\tau ,S)$%
\begin{equation}
\frac{\partial Z^{\varepsilon }}{\partial \tau }=\frac{1}{2}\overline{\sigma
^{2}}S^{2}\frac{\partial ^{2}Z^{\varepsilon }}{\partial S^{2}}+\frac{1}{2}%
v\left( \frac{\tau }{\varepsilon }\right) S^{2}\frac{\partial
^{2}Z^{\varepsilon }}{\partial S^{2}}+rS\frac{\partial Z^{\varepsilon }}{%
\partial S}-rZ^{\varepsilon }+\frac{1}{2}\varepsilon ^{-1/2+d}v\left( \frac{%
\tau }{\varepsilon }\right) S^{2}\frac{\partial ^{2}\overline{C}}{\partial
S^{2}}.  \label{error}
\end{equation}
Our objective now is to find the asymptotic limit of $Z^{\varepsilon }(\tau
,S)$ as $\varepsilon \rightarrow 0.$ The equation (\ref{error}) involves two
stochastic terms:
\begin{equation}
\frac{1}{2}v\left( \frac{\tau }{\varepsilon }\right) S^{2}\frac{\partial
^{2}Z^{\varepsilon }}{\partial S^{2}},\;\;\;\;\;\;\frac{1}{2}\varepsilon
^{-1/2+d}v\left( \frac{\tau }{\varepsilon }\right) S^{2}\frac{\partial ^{2}%
\overline{C}}{\partial S^{2}}.
\end{equation}
Ergodic theory implies that the first term in its integral form converges to
zero as $\varepsilon \rightarrow 0,$ while the second term converges weakly
to
\begin{equation}
\frac{1}{2}D_{d}^{1/2}\xi _{d}(\tau )S^{2}\frac{\partial ^{2}\overline{C}}{%
\partial S^{2}}\;\;as\text{ }\;\;\varepsilon \rightarrow 0,  \label{conver}
\end{equation}
where $\xi _{d}(\tau )$ is the fractional Gaussian white noise
\begin{equation}
\xi _{d}(\tau )=\frac{dB_{d}}{d\tau }  \label{noise}
\end{equation}
with the covariance \cite{Hu,Duncan}
\begin{equation}
E\{\xi _{d}(\tau )\xi _{d}(s)\}=d(2d+1)|\tau -s|^{2d-1}.  \label{covfr}
\end{equation}
Here, $B_{d}$ is the fractional Brownian motion (\ref{Fractional}) and $%
D_{d} $ is given by
\begin{equation}
D_{d}=\frac{2c}{(1-\alpha )(2-\alpha )},\;\;0<\alpha <1  \label{D}
\end{equation}
(derivation of $D_{d}$ can be found in the Appendix A). The limit (\ref
{conver}) follows from \cite{Gira}
\begin{equation}
\varepsilon ^{-1/2+d}\int_{0}^{\tau }v\left( \frac{s}{\varepsilon }\right)
ds\rightarrow D_{d}^{1/2}B_{d}(\tau )\;\;as\text{ }\;\;\varepsilon
\rightarrow 0.
\end{equation}
Thus the random field $Z^{\varepsilon }(\tau ,S)$ converges weakly to $%
Z(\tau ,S)$ which obeys the asymptotic equation
\begin{equation}
\frac{\partial Z}{\partial \tau }=\frac{1}{2}\overline{\sigma ^{2}}S^{2}%
\frac{\partial ^{2}Z}{\partial S^{2}}+rS\frac{\partial Z}{\partial S}-rZ+%
\frac{1}{2}D_{d}^{1/2}S^{2}\frac{\partial ^{2}\overline{C}}{\partial S^{2}}%
\xi _{d}(\tau ).  \label{limitequation}
\end{equation}
with the initial condition $Z(0,S)=0$.

In this paper we present only a heuristic derivation of the limiting
equation (\ref{limitequation}). The rigorous derivation should involve the
exact renormalization technique together with stochastic averaging and the
central limit theorem for stochastic processes \cite{Avellaneda,Watanabe}.
It should be noted that Eq. (\ref{limitequation}) can be also rewritten in
terms of the Black-Scholes operator
\begin{equation}
L_{BS}\left[ Z\right] =\frac{1}{2}\overline{\sigma ^{2}}S^{2}\frac{\partial
^{2}Z}{\partial S^{2}}+rS\frac{\partial Z}{\partial S}-rZ
\end{equation}
and the fractional Brownian motion $B_{d},$ as follows:
\begin{equation}
dZ=L_{BS}\left[ Z\right] dt+\frac{1}{2}D_{d}^{1/2}S^{2}\frac{\partial ^{2}%
\overline{C}}{\partial S^{2}}dB_{d}.
\end{equation}
The advantage of the asymptotic equation (\ref{limitequation}) is that it
can be solved in terms of the classical Green's function, $G(S,S_{1},\tau
,\tau _{1}),$ for the Black-Scholes equation \cite{Wilmott}. It follow from (%
\ref{limitequation}) that \cite{C} \
\begin{equation}
Z(\tau ,S)=\frac{1}{2}D_{d}^{1/2}\int_{0}^{\tau }\int_{0}^{\infty
}G(S,S_{1},\tau ,\tau _{1})S_{1}^{2}\frac{\partial ^{2}\overline{C}}{%
\partial S^{2}}(\tau _{1},S_{1})\xi _{d}(\tau _{1})dS_{1}d\tau _{1},
\label{solution}
\end{equation}
where $G(S,S_{1},\tau ,\tau _{1})$ is
\begin{equation}
G(S,S_{1},\tau ,\tau _{1})=\frac{e^{-r(\tau -\tau _{1})}}{S_{1}\sqrt{2\pi
\overline{\sigma ^{2}}(\tau -\tau _{1})}}\exp \left( -\frac{[\ln
(S/S_{1})+(r-\frac{\overline{\sigma ^{2}}}{2})(\tau -\tau _{1})]^{2}}{2%
\overline{\sigma ^{2}}(\tau -\tau _{1})}\right) .  \label{Green}
\end{equation}
The variance
\begin{equation}
V(\tau ,S)=E\{Z^{2}(\tau ,S)\}
\end{equation}
can be regarded as a measure of volatility risk. It can be easily found from
(\ref{solution}) that
\begin{gather}
V(\tau ,S)=\frac{1}{4}D_{d}\int_{0}^{\tau }\int_{0}^{\tau }\int_{0}^{\infty
}\int_{0}^{\infty }G(S,S_{1},\tau ,\tau _{1})G(S,S_{2},\tau ,\tau _{2})
\label{Zcov} \\
\times S_{1}{}^{2}S_{2}{}^{2}\frac{\partial ^{2}\overline{C}}{\partial S^{2}}%
\left( \tau _{1},S_{1}\right) \frac{\partial ^{2}\overline{C}}{\partial S^{2}%
}(\tau _{2},S_{2})E\{\xi _{d}(\tau _{1})\xi _{d}(\tau
_{2})\}dS_{1}dS_{2}d\tau _{1}d\tau _{2},  \notag
\end{gather}
where
\begin{equation}
E\{\xi _{d}(\tau _{1})\xi _{d}(\tau _{2})\}=d(2d+1)|\tau _{1}-\tau
_{2}|^{2d-1}.  \label{cov2}
\end{equation}
It should be noted that despite the fact that $\xi _{d}(\tau )$ converges to
the Gaussian white noise as $d\rightarrow 0$ and
\begin{equation}
\lim_{d\rightarrow 0}E\{\xi _{d}(\tau _{1})\xi _{d}(\tau _{2})\}=\delta
(\tau _{1}-\tau _{2}),
\end{equation}
we cannot consider the short memory case ($d=0)$ for (\ref{Zcov}). The
problem is that in the limit $d\rightarrow 0$ ($D_{d}\rightarrow \infty )$
we have logarithmic divergence of time integrals (see Appendix A) . Recall
that in the Gaussian white noise case the variance $V(\tau ,S)=$ $%
E\{Z^{2}(\tau ,S)\}$ has the following expression \cite{Sircar2}
\begin{equation}
V(\tau ,S)=\frac{D}{4}\int_{0}^{\tau }\left( \int_{0}^{\infty
}G(S,S_{1},\tau ,\tau _{1})S_{1}^{2}\frac{\partial ^{2}\overline{C}}{%
\partial S_{1}^{2}}dS_{1}\right) ^{2}d\tau _{1}.  \label{zero}
\end{equation}

It should be noted that the main result concerning the measure of
volatility risk (\ref{Zcov}) is very robust. The main reason for
this is that for the random volatility $\sigma ^{2}(t)$ we use a
quite general stationary random process (\ref{covar1}) without
assuming the exact distributions for it, like Gaussian, Poisson,
etc. It is very important because there is empirical evidence that
the volatility is not Gaussian and its probability density
function can have power-law tails \cite{last}. Of course after
rescaling the central limit theorem insures in the asymptotic
limit $\varepsilon \rightarrow 0$ effective volatility becomes
Gaussian. It follows from (\ref {covar1}) that the value of the
covariance $g(s)$ at $s=0$ is finite. One can consider the
situation when this is not the case. Then we should apply a
different scaling procedure which will lead to a stable
distribution with the power-law tails in the asymptotic limit
\cite{Samor}.

\section{Numerical Results}

In this section we present numerical results for the variance of $%
C^{\varepsilon }(\tau ,S)$
\begin{equation}
E\left\{\left( C^{\varepsilon }(\tau ,S)-\overline{C}(\tau ,S)\right)
^{2}\right\} =\varepsilon ^{1-2d}V(\tau ,S)  \label{measure}
\end{equation}
for the different values of $d$. Assume that the exercise price of the call
option is $\$50$, the risk-free interest rate is $5\%$ p.a., and the average
volatility is $20\%$ p.a., that is, $K=50,$ $r=0.05,$ $\overline{\sigma ^{2}}%
=0.04.$ Figure 1 shows the graph of the variance $E\left\{ \left(
C^{\varepsilon }(\tau ,S)-\overline{C}(\tau ,S)\right) ^{2}\right\} $plotted
against the stock price $S$ for $\tau =0.5$ , $\varepsilon =0.1$ and $d=0.01$%
, $0.1$, $0.3$, and $0.45$ respectively ($c=1$).

From Figure 1 we can see that for at-the-money call option there is maximum
uncertainty since all graphs have a maximum near $S=K,$ where $K$ is the
strike price. As for deep in- and out-the-money option there is almost zero
uncertainty. It should be noted that the strength of the memory effect is
determined by the value of the parameter $d.$ One can see from Figure 1 that
the increase in $d$ leads to an increase in the variance and therefore in
the risk value. If we assume as in, \cite{Sircar2}, that the writer sells
the option for
\begin{equation}
\overline{C}(\tau ,S)+2\varepsilon ^{1/2-d}V^{1/2}(\tau ,S)  \label{option}
\end{equation}
then the increase in parameter the $d$ leads to the increase in option
price. In Figure 2, we plot 3-dimensional graphs of the variance $E\left\{
\left( C^{\varepsilon }(\tau ,S)-\overline{C}(\tau ,S)\right) ^{2}\right\} $%
against the stock price $S$ and time to maturity $\tau $ for $d=0.3$ and $%
d=0.1.$ Note that the variance increases with the time to maturity. It
should be noted that these results are similar to those obtained in \cite
{Aurel}. It was found that the residual risk vanishes for both deep
in-the-money options and deep out-of-the money options, but increases
sharply for at-the-money option.

\section{ Conclusion}

In this paper we presented a stochastic model that accounts for the effects
of a long-memory in volatility on option pricing. We started with the random
Black-Scholes equation with the volatility parameter being the stationary
random process with long-range dependence. We represented the option price
as a sum of classical Black-Scholes price and random deviation describing
the risk from the random volatility. By using the renormalization procedure
we derived the asymptotic equation for this deviation. By using Green's
function methods we solved this equation and found the asymptotic pricing
bands for options.

From numerical calculations of the variance of the random option price we
found that the maximum deviation from the classical Black-Scholes price
takes place for at-the-money option, while the risk due to random volatility
vanishes both for deep-in and out-of-the-money options. We also found that
the increase in the strength of the memory effect leads to an increase in
the variance and therefore in the effective option price.

\renewcommand{\theequation}{A-\arabic{equation}} \setcounter{equation}{0}

\section*{Appendix A}

\textbf{Weak convergence to fractional Brownian motion}

\ The purpose of this Appendix is to derive the expressions for the
parameters $D_{d}$ and $d$ by using the autocorrelation function
\begin{equation}
g(s)=\;E\{v(\tau )v(\tau +s)\}=\frac{c}{(s_{0}+s)^{\alpha }},\;\;\;0<\alpha
<1.  \label{covA}
\end{equation}
Let us define the process $x^{\varepsilon }(\tau )$ as follows:
\begin{equation}
x^{\varepsilon }(\tau )=\varepsilon ^{-\frac{1}{2}+d}\int_{0}^{\tau }v\left(
\frac{s}{\varepsilon }\right) ds=\varepsilon ^{\frac{1}{2}+d}\int_{0}^{\frac{%
\tau }{\varepsilon }}v(s)ds.  \label{new}
\end{equation}
It converges weakly to the scaled fractional Brownian motion $%
D_{d}^{1/2}B_{d}$ as $\varepsilon \rightarrow 0.$ To derive expressions for $%
D_{d}$ and $d$, we show that
\begin{equation}
\lim_{\varepsilon \rightarrow 0}E\{[x^{\varepsilon }(\tau )]^{2}\}=D_{d}\tau
^{2d+1},
\end{equation}
where
\begin{equation}
D_{d}=\frac{2c}{(1-\alpha )(2-\alpha )},\;\;d=\frac{1-\alpha }{2}.
\end{equation}
It follows from (\ref{covA}) and (\ref{new}) that
\begin{equation}
E\{[x^{\varepsilon }(\tau )]^{2}\}=\varepsilon ^{1+2d}\int_{0}^{\frac{\tau }{%
\varepsilon }}\int_{0}^{\frac{\tau }{\varepsilon }%
}g(s_{1}-s_{2})ds_{1}ds_{2}.  \label{div}
\end{equation}
By using the well-known formula for stationary process $v(s)$ with zero mean
and the covariance function $g(s)$
\begin{equation}
\int_{0}^{\tau }\int_{0}^{\tau
}E\{v(s_{1})v(s_{2})\}ds_{1}ds_{2}=\int_{0}^{\tau }\int_{0}^{\tau
}g(s_{1}-s_{2})ds_{1}ds_{2}=2\tau \int_{0}^{\tau }g(s)ds-2\int_{0}^{\tau
}sg(s)ds,
\end{equation}
one finds that
\begin{equation}
E\{[x^{\varepsilon }(\tau )]^{2}\}=2\varepsilon ^{1+2d}\left( \frac{\tau }{%
\varepsilon }\int_{0}^{\frac{\tau }{\varepsilon }}g(s)ds-\int_{0}^{\frac{%
\tau }{\varepsilon }}sg(s)ds\right) .  \label{seven}
\end{equation}
In the long memory case ( $0<\alpha <1)$ both integrals
\begin{equation}
\int_{0}^{\frac{\tau }{\varepsilon }}g(s)ds,\;\;\;\int_{0}^{\frac{\tau }{%
\varepsilon }}sg(s)ds
\end{equation}
diverge as $\varepsilon \rightarrow 0.$ That is why we need here the
appropriate scaling factor $\varepsilon ^{1+2d}$ to ensure that $%
E\{x^{\varepsilon }(\tau )\}^{2}$ is finite for the fixed value of $\tau .$
By using (\ref{covA}) we find
\begin{equation}
\int_{0}^{\frac{\tau }{\varepsilon }}g(s)ds=\frac{c}{\left( 1-\alpha \right)
}\allowbreak \left[ \left( \frac{s_{0}\varepsilon +\tau }{\varepsilon }%
\right) ^{-\alpha }s_{0}+\left( \frac{s_{0}\varepsilon +\tau }{\varepsilon }%
\right) ^{-\alpha }\tau \varepsilon ^{-1}-s_{0}^{1-\alpha }\right]
\end{equation}
and
\begin{eqnarray}
\int_{0}^{\frac{\tau }{\varepsilon }}sg(s)ds &=&-\frac{c}{(1-\alpha
)(2-\alpha )}\left( \frac{s_{0}\varepsilon +\tau }{\varepsilon }\right)
^{-\alpha } \\
&&\left[ \tau ^{2}\varepsilon ^{-2}\alpha +s_{0}\alpha \tau
\varepsilon ^{-1}-\tau ^{2}\varepsilon ^{-2}+s_{0}^{2}-\left(
\frac{s_{0}\varepsilon +\tau }{\varepsilon }\right) ^{\alpha
}s_{0}^{2-\alpha }\right].
\end{eqnarray}
If we set
\begin{equation}
d=\frac{1-\alpha }{2},
\end{equation}
we find from (\ref{seven}) that
\begin{equation}
\lim_{\varepsilon \rightarrow 0}E\{x^{\varepsilon }(\tau )\}^{2}=\frac{%
2c\tau ^{2-\alpha }}{(1-\alpha )(2-\alpha )}
\end{equation}
and therefore
\begin{equation}
D_{d}=\frac{2c}{(1-\alpha )(2-\alpha )}.
\end{equation}

\newpage

\begin{figure}[p]
\centering
\includegraphics[scale=0.8]{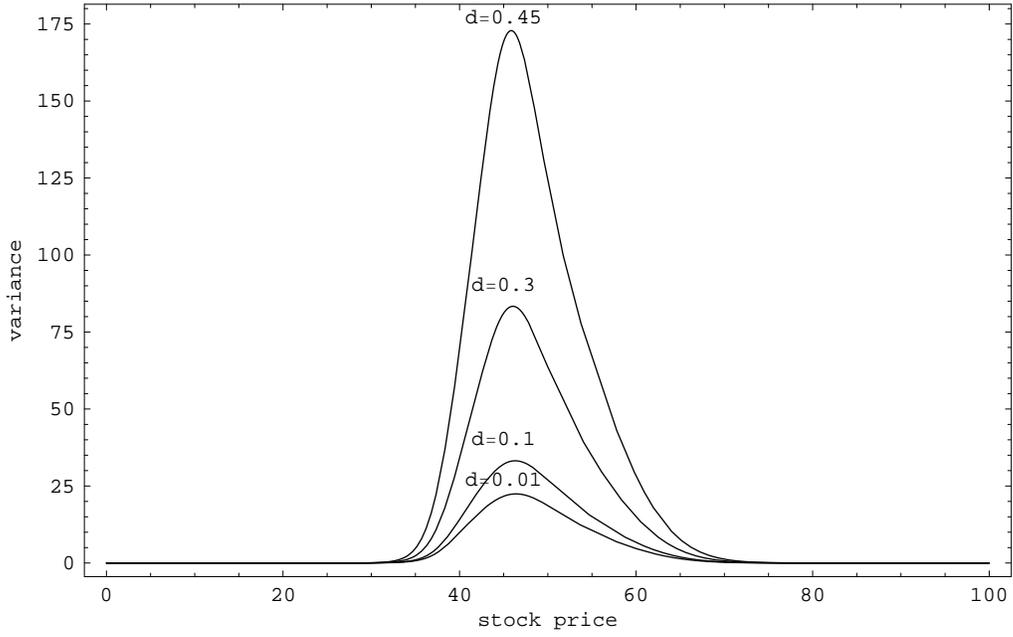} ~
\caption{Plot of variance, $E{\left\{\left( C^{\protect\varepsilon }(\protect%
\tau ,S)-\overline{C}(\protect\tau ,S)\right) ^{2}\right\}}$
against stock
price, $S$ for $d=0.01$, $d=0.1$, $d=0.3$ and $d=0.45$ with $K=50$ and $%
\protect\tau =0.5.$ } \label{figure 1}
\end{figure}
\begin{figure}[p]
\centering
\includegraphics[scale=0.6]{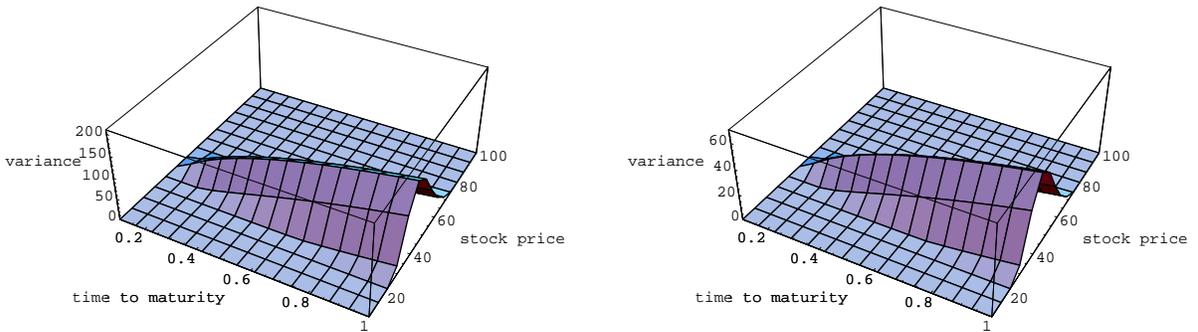} ~
\caption{3D-plot of variance, $E{\left\{\left( C^{\protect\varepsilon }(%
\protect\tau ,S)-\overline{C}(\protect\tau ,S)\right)
^{2}\right\}}$ against stock price, $S$ and time to maturity,
$\protect\tau $ for $d=0.3$ (left) and $d=0.1$ (right) with $K=50$
and $\protect\tau =0.5.$} \label{figure 2}
\end{figure}

\end{document}